\documentstyle[aps,preprint]{revtex}
\usepackage{psfig}

\begin{document}

\title{Designability of lattice model heteropolymers}
\author{G. Tiana$^1$, R. A. Broglia$^{1,2}$ and D. Provasi$^{1}$ \\
        $^1$ Dipartimento di Fisica, Universit\'a di Milano e\\
	INFN Sezione di Milano, via Celoria 16, 20133 Milano, Italy\\
	$^2$ The Niels Bohr Institute, Blegdamsvej 17, \\
        2100 Copenhagen, Denmark}
\date{\today}
\maketitle

\begin{abstract}
Protein folds are highly designable, in the sense that many sequences fold to
the same conformation. In the present work we derive an expression 
for the designability in a  20 letter
lattice model of proteins which, relying only on the Central Limit Theorem, has a
generality which goes beyond the simple model used in its derivation. This expression displays an
exponential dependence on the energy of the optimal sequence folding on the given
conformation measured with respect to the lowest energy of the conformational dissimilar structures, 
energy difference which constitutes the only parameter controlling designability.
Accordingly, the designability of a native conformation 
is intimately connected to the stability of the sequences folding to them. 
\end{abstract}

\section{Introduction}

Even a quick look at the set of known proteins (PDB database) reveals a striking feature.
While there are tens of thousands of protein sequences, they only assume some thousands
folds. In other words, proteins are highly designable. This concept can be
quantified by measuring the number of sequences that fold uniquely into a particular
structure.

With the use of a simple 20 letter lattice model \cite{sh_prl,sh_gap,r3,r4} 
of protein folding it has been shown
\cite{jchemphys3} that the
whole issue of estimating the number $n$ of sequences
that fold to the same conformation is reduced to enumerate how many of them have native energy
lying below a threshold $E_c$, 
the energy which any sequence with the same composition displays
in the conformation structurally dissimilar to the native conformation \cite{sh_gap}.

The aim of the present paper is to
provide a reliable, analytic expression for $n$, which we shall show
increases exponentially with the gap $\delta$ between the native energy of the optimal sequence $E_n$ and
the threshold energy $E_c$. This functional form is found to be universal, as it emerges
from the Central Limit Theorem. We have furthermore found that while
the parameters defining $n$ depend on
the interaction matrix, they are independent of the particular choice made for
the native structure or for the
optimal sequence.

In Sections 2 and 3 we briefly review the 20 letters lattice model of proteins in general
and the question of protein designability in particular. The quantitative, analytic answer
to the question of how many mutations a designed protein tolerates is given in Section
4. The conclusions are collected in Section 5.

\section{Lattice Models}

A useful theoretical approach to study protein folding is provided by a simplified lattice
model, where the protein is a string of beads that is arranged on a cubic
lattice \cite{go,dill,ss1}. The configurational energy of a chain of N monomers is given by
\begin{equation}
E=\frac{1}{2}\sum^N_{i,j}U_{m(i),m(j)}\Delta(|\vec{r}_i-\vec{r}_j|),
\end{equation}
where $U_{m(i),m(j)}$ is the effective interaction potential between monomers
$m(i)$ and $m(j)$, $\vec{r}_i$ and $\vec{r}_j$ denote their lattice positions
and $\Delta(x)$ is the contact function. In Eq. (1) the pairwise interaction is
different from zero when $i$ and $j$ occupy nearest--neighbour sites, i.e.,
$\Delta(a)=1$ and $\Delta(na)=0$ for $n\geq 2$, where $a$ indicates the step
length of the lattice. In addition to these interactions, it is assumed that
on--site repulsive forces prevent two amino acids to occupy the same site
simultaneously, so that $\Delta(0)=\infty$. 

We shall consider throughout a 20--letters
representation of protein sequence where $U$ is a $20\times 20$ matrix. A
possible realization of this matrix is given in ref. \cite{mj}
(Table 5 and 6), where it was derived from frequencies of contacts between different
amino acids in protein structures. The employment of a $20\times 20$ matrix ensures
that the threshold energy $E_c$ is well defined, depending only on the
interaction matrix elements and on the composition of the protein in terms of amino acids.
The model we study here is a generic
heteropolymer model which has been shown to reproduce important generic features
of protein folding thermodynamics and kinetics, independent on the particular
potential chosen \cite{sh_b1,sh_b2}. However, in using such an approach, one should
keep in mind that the labelling of amino acids (spherical beads all of the same
size and with no side chain) is generic too and may be no obvious relation
between those labels and labels for real amino acids.

Good--folder sequences are characterized by a large gap $\delta=E_c-E_n$
(compared to the standard deviation $\sigma$ of the contact energies) between
the energy of the designed sequence in the native conformation $E_n$, and the lowest
energy of the conformations structurally dissimilar to the
native conformation \cite{sh_prl,sh_gap,r3,r4,jchemphys3}. 
In other words, good folders are associated with a normalized gap
$\xi=\delta/\sigma\gg 1$, quantity closely related to the z--score \cite{bowie}.
For example, the 36mer sequence listed in the caption to Fig. 1 and called S$_{36}$ in the literature
\cite{jchemphys2,n1,sh_nucleus,n2,n3,jchemphys1}, designed by minimizing 
the energy in the target (native) conformation with respect
to amino acid sequence for fixed composition has, in the units considered here 
($RT_{room}=0.6 \;kcal/mol$ \cite{mj}), an energy gap $\delta=2.5$ and thus a sufficiently large value of
$\xi$ ($=2.5/0.3\approx 8.33$) so as to ensure fast folding. In fact,
Monte Carlo simulations carried out at the temperature $T=0.28$ of 3000 36mers
with energies, in the native conformation, lying inside the gap fold in times $\leq
7\times 10^7$ MC steps \cite{jchemphys3} (For caveats see ref. \cite{note_new}). In particular S$_{36}$ folds in $6.5\times 10^6$
MC steps. 

It has been also shown that most of the thermodynamical \cite{jchemphys1} and dynamical
\cite{jchemphys3,jchemphys2}
behaviour of designed proteins is controlled by only $5-10\%$ of the sites. As a consequence, making
mutations in these sites, which are called "hot" in ref. \cite{jchemphys1}, one destroys,
as a rule, the
ability the protein has to fold (denaturation). On the other hand, the effects of substitutions in any other
site (that can be regarded as "cold") are small, leading to neutral mutations.

\section{Designability with 20 letters models}

While twenty letters heteropolymers capture the essential components of
real proteins, it is
hardly possible to enumerate all sequences which have a given conformation as their
non--degenerate ground state. Accordingly, it is not possible to calculate the
exact designability of protein conformations.
To bypass this problem, we shall determine designability
from energetic considerations, using a strategy which relies on 
the fact that all sequences 
which have an energy lower than $E_c$ fold on short call, in any case in times 
which are much shorter than that associated with the random search \cite{prl}.

Any sequence of a given length $N$ (e.g. $N=36$) 
can be obtained making $m\leq N$
mutations (i.e. substitution of a amino acid in a given site with a different one)
in the minimum energy sequence (e.g. S$_{36}$ in the case of Fig. 1(a) ).
Consequently, the designability of a conformation 
can be found starting from the minimum energy sequence, 
counting how many mutations lay within the gap
$\delta=E_c-E_n$.  If $\Delta E$ is the change in the energy of the native state
produced by a mutation, $p_m(\Delta E)$ the energy distribution probability
associated with $m$ mutations 
and $n^{tot}_m$  the total number of sequences that can be produced by introducing $m$
mutations in the minimum energy sequence, designability can be defined as
\begin{eqnarray}
\label{def_desig}
n & = & \sum_{m=1}^N n_m ,\\
n_m & = & n^{tot}_m\int_{0}^{\delta} p_m(\Delta E)\; d(\Delta E).
\end{eqnarray}

So far, we have done nothing more than expressing the problem in another way,
since to know the spectrum of mutation energies of the optimal sequence
one has again to enumerate all sequences. In fact, it looks like as if we
have made things even worse, in that now one has to find the optimal
sequence, which is a non--trivial problem, and also has to ensure that
$E_c$ does not change with mutations.

We shall show in the following that the distribution 
of mutation energies does not depend on the
particular structure or on the particular sequence chosen
(provided that  $E\ll E_c$) nor on the contact energy matrix used to design the
protein, but only on its composition
and on the number of contacts (or the length, if it is fully compact).
This observation leaves room to approximations.
In fact, if one is able to find an approximate expression for $p(\Delta E)$, 
such an expression will hold for all  model proteins of the same
length. Furthermore, the knowledge of the sequence associated with the 
global minimum of energy $E_{n}$ is not necessary (because all sequences
have the same spectrum of mutations), only the value of
$E_n$ is required. Consequently, even if the optimal 
sequence cannot be known without a full enumeration of all sequences,
it is allowed to use any other sequence with energy $E\approx E_n$,
introducing in this way only an error in the integration boundaries (and not on the function
to be integrated). It is then possible to calculate designability of
a structure
from Eq. (2--3) using an approximate distribution $p(\Delta E)$
and an approximate value of $\delta$. 

The most conservative
way to calculate the number of sequences which fold to a conformation is
then to use a distribution $p_m(\Delta E)$ found only by swappings between the residues of
the optimal sequence, as in such a way the composition is conserved
and $E_c$ does not change. On the other hand, since there are also sequences with different
composition folding to the same conformation, one is also forced, in principle,
to calculate
the number $n$ associated with pointlike mutations. The values found from swapping of amino
acids
and from pointlike mutations can be viewed as the lower and the upper limit to
designability, respectively.

In Figs. 2(a) and 2(b) we display the unnormalized energy 
distribution probabilities associated with two composition--conserving and with
two pointlike mutations of S$_{36}$ (the integral of these distributions being
the total number of sequences). Each of
these curves can be well fitted by the sum of 
two Gaussians, whose means are $\overline{\Delta E_2}=1.2$ and 
$\overline{\Delta E_2}=3.0$ (Fig. 2(a), composition conserving case) and
$\overline{\Delta E_2}=1.1$ and $\overline{\Delta E_2}=3.6$ 
(Fig. 2(b), pointlike mutations case). Standard deviations are 
$\sigma_{2}=0.7$
and $\sigma_{2}=1.0$ (Fig. 2(a) ) and $\sigma_{2}=0.7$ and
$\sigma_{2}=1.1$ (Fig. 2(b) ). 
The behaviour of these two distributions seems very alike, except for the fact
that the area below the composition--conserving curve is much smaller than that
below the pointlike mutations curve. This is because much fewer mutations can
be made in the first than in the second case and, consequently, the associated
Gaussian behaviour is less well defined. 

The overall structure of the curves shown in Fig. 2
can be understood from the fact that the average value of $\Delta E$ for "cold" sites
is $0.65$ and for "hot" sites is $2.87$ 
\cite{jchemphys1}. Accordingly, the low--energy peak can be associated
with two mutations in "cold" sites, while the high--energy peak can be associated with
a mutation in a "cold" site and a mutation in a "hot" site. The contribution
from mutations in two "hot" sites leads to an enhancement of the high energy tail of the
curve. Concerning the Gaussian behaviour, we note that the energies associated with
the 19 possible mutations on a given "cold" site are uncorrelated. In other words,
one has to pay an energy $\overline{\Delta E_2}/2\approx 0.6$ (concerning the factor $1/2$ one
is reminded of the fact that $\overline{\Delta E_2}$ gets contributions from two mutations) 
to remove the wild--type
residue, reflecting the fact that it has been optimized. Second, one has to introduce a new
residue in the {\it niche} left by the wild--type residue. The Gaussian shape of the
distribution suggests that the {\it niche} is neutral with respect 
to the new residue and that the new interactions
are merely random. To be more precise, the change in energy $\Delta E$ upon mutations is the
difference between the energy needed to remove the original residue (which is roughly
constant and assumes two different values for cold and for hot sites) 
and the sum of a number of contact energies associated with the new residue, energies 
which can be considered as random numbers. Being the sum of random 
numbers, the energy associated with the new
residue is forced to respect the 
Central Limit Theorem, and consequently its distribution approaches a Gaussian
function. Of course, an exact Gaussian distribution could be reached only if the
number of nearest neighbours of each site were infinite (while in a cubic
lattice this is, at most, five). On the other hand, the fact that  
$p_m(\Delta E)$ can be accurately fitted by Gaussian distributions (cf.,
e.g., Fig. 2(b) ) testifies to the fact that we are not far from the conditions
in which the Central Limit Theorem holds.

While the hypothesis that "cold" mutations give raise to Gaussian--like peaks is
quite grounded, due to the uncorrelateness of the energy contributions of
"cold" sites, it is unlikely that the Central Limit Theorem works properly for
"hot" sites, whose energy contributions are correlated \cite{jchemphys3}. In
order to calculate the degree of designability of a protein conformation, we
only need to know the contributions from cold sites and, consequently, we don
not need to better characterize the peaks associated to "hot" sites. 

We have found that the distribution of mutation energies are rather universal functions.
Examples of such a behaviour are shown in Fig. 3, where
2--pointlikes mutations spectra $p_2(\Delta E)$ associated with low energy 36mer sequences 
optimized (making use of the MJ elements of Table 6) on three
different conformation (cf. Figs. 1(a)--1(c) ) and with three sequences designed on the same
conformation (Fig. 1(a) ) are displayed. Similar results have been obtained for chains of
different lengths. Furthermore, using different 
$20\times 20$ interaction matrices lead to the same Gaussian behaviour of $p_2(\Delta E)$,
although the mean values and the standard deviations are different. This is again a
consequence of the Central Limit theorem. This can be seen from Fig. 4, where
we display the function  $p_2(\Delta E)$ associated with two pointlike 
mutations on S$_{36}$ (cf. Fig. 1), but
making use this time of
the interaction matrix elements listed in Table 5 of ref. \cite{mj}. Because  
the average change in energy upon mutations in cold sites is zero, 
while that in hot sites is $0.35$, it is easy to identify the peaks
associated with two cold mutations ($\overline{\Delta E_2}=0$ and $\sigma_2=0.34$), with
one cold and one hot mutations ($\overline{\Delta E_2}=0.35$ and $\sigma_2=0.02$), 
and with two hot mutations ($\overline{\Delta E_2}=0.70$ and $\sigma_2=0.22$).

Summing up, the function $p_2(\Delta E)$ associated with chains of different length and
sequence as well as designed on different native conformations overlap quite nicely,
suggesting that the spectrum of both composition conserving and non--conserving 
mutations is universal. On the other hand, the actual value of  $\overline{\Delta E_2}$ and
$\sigma_2$ characterizing the different peaks of the
energy distribution probability depend on the matrix used to describe the
contact energies among the amino acids.

The univerality of the energy distribution probability  
is in agreement with the interpretation of the main peaks of the spectrum of mutations of a designed
protein
in term of "cold" and "hot" sites. In fact, the properties of the "hot" sites 
are rather homogeneous,
their contribution to the mutation spectrum being universal. 
Assuming furthermore that the interactions associated to 
the  mutations in "cold" sites are random,
the resulting energy distribution is Gaussian 
(and so universal), and its standard
deviation depends only on the interaction matrix, while its average value depends 
on the degree of
optimization of the wild--type monomer. This, in turn, can be approximated by the 
degree
of optimization of the whole chain (measured by the energy gap $\delta$)
divided by its length, a quantity which is essentially  constant
for long chains \cite{note3} (for example, in the case of S$_{36}$ this number is
$2.5/36=0.07$).

\section{How many good folders?}

The basic idea to calculate the designability of a model protein, as we have discussed above,
is to find a simple approximation to the universal distribution of energies associated
with mutations onto the optimal sequence, integrate this distribution up to the gap $\delta$ and
normalize this result to the total number of mutations that one can make (cf. Eqs. (2--3) ). As a
consequence of this, designability turns out to depend only on the 
length of the protein (through the total number of mutations) and on the gap $\delta$.

In order to calculate $n$ using Eq. (2--3) we have first to know the total number $n^{tot}_m$
of sequences that can be obtained by making $m$ sequence--conserving mutations 
(swappings) in the optimal sequence.
This number can be obtained by counting 
the number of ways one can select $m$ sites, multiplied by
the number of permutations of these sites which move all the $m$ residues. That is
\begin{equation}
n^{tot}_{m}={N \choose m}P_0(m),
\end{equation}
where $P_i(m)$ is the number of ways one can permute m sites in
such a way that only $i$ positions are kept fixed. From the relation
\begin{equation}
m! =  P_0(m) + P_1(m) + P_2(m) + ... + P_{m-2}(m) +1
\end{equation}
it is possible to extract the expression for $P_0$, 
\begin{equation}
P_0(m)=m!-\sum_{k=1}^{m-2}{m \choose k} P_0(m-k) -1.
\end{equation}
For large $m$, one can use the
Stirling approximation for the factorials in Eqs. (4--6), and keep only the largest
exponent term in the sum (saddle point approximation), obtaining 
$n^{tot}_m\approx\exp(\alpha m)$. The constant $\alpha$ 
can be determined from the relation $n^{tot}_N=e^{\alpha N}=N!$, which
for $N=36$ leads to $\alpha=2.66$. 
 
To proceed further in the calculation of $n$, one needs to find a simple approximation 
to $p_m(\Delta E)$.
For this purpose, we shall express the energy distribution of an arbitrary number
of mutations as a convolution of functions $p_2(\Delta E)$ associated with the swapping of two
amino acids. The validity of this approximation rests on the ansatz 
that every couple of mutations affect the energy of the native state
independently of the other couple of mutations. This approximation is
expected to work also for large values of
$m$, where the probability of mutating neighbouring sites is not negligible, because the
contact energy associated with the mutated residues are in any case random quantities with
average zero (cf. the discussion in the previous section).
Within this scenario, the
number of folding sequences displaying $2m$ mutations and whose energy in the native
conformation lies inside the energy gap can be written as
\begin{eqnarray}
n_{2m}\approx n^{tot}_{2m}  \int_{0}^{\delta}dE\;
\int_{-\infty}^{+\infty} d\Delta E_1\;d\Delta E_2...d\Delta E_{m-1} \times \nonumber \\
\times p_2(\Delta E_1)\;p_2(\Delta E_2)\;...\;p_2(\Delta E_{m-1})\;p_2(\Delta E-\Delta E_1-
\Delta E_2-...-\Delta E_{m-1}). 
\end{eqnarray}
Making use of the energy distribution probability associated with an amino acid swapping
(composition conserving mutations) or with two point--like mutations (composition
non--conserving mutations) one obtains the lower and the upper limit of the designability of a
conformation. 

In what follows we shall essentially discuss the case of composition conserving
mutations. If $\delta$ is lower than the peak associated
with mutations in "hot" sites (as in the case of the sequence
S$_{36}$ where $\delta=2.5$, cf. Fig. 2),
one should convolute only the peak of $p_2(\Delta E)$ 
associated with mutations in "cold" sites \cite{note4}. 
Exploiting the fact that the convolution of $m$ Gaussian
distributions, of the form $\exp((\Delta E-\overline{\Delta E}_2)^2/2(\sigma_2)^2)$ 
is a Gaussian distribution
with average $\overline{\Delta E}_{2m}=m\overline{\Delta E}_2$ and standard deviation
$\sigma_{2m}=m^{1/2}\sigma_2$, it is possible to write
\begin{eqnarray}
\label{des_final}
n_{2m}\approx  n^{tot}_{2m} (2\pi m^2\sigma_2^2)^{-1/2}
\exp\left( \frac{-(\overline{\Delta E}_2)^2}{2(\sigma_2)^2} m \right)\times \nonumber \\
\times\int_{0}^{\delta}d\Delta E\; \exp\left(-\frac{\Delta E^2}{2m(\sigma_2)^2}+
\frac{\overline{\Delta E}_2 \Delta E}{2(\sigma_2)^2}\right).
\label{eq5}
\end{eqnarray}
For
$m\gg\delta/(2\sigma_2)^{1/2}$ (in the case of S$_{36}$ this condition means $m\gg 2$) one can
neglect
the first exponential factor in the integral, in which case the integration can be carried out
analytically, leading to 
\begin{equation}
n_{m}\approx n^{tot}_{m} \;(\pi m^2\sigma_2^2/2)^{-1/2}\;
\exp\left(-\frac{\overline{\Delta E}_2^2}
{4(\sigma_2)^2}m\right)\;\frac{2(\sigma_2)^2}{\overline{E}_2}
\left(\exp\left[\frac{\overline{\Delta E}_2}
{2(\sigma_2)^2}\delta\right]-1\right)
\label{eq6}
\end{equation}
where the substitution $2m\rightarrow m$ has been made. 
This equation tells us  that designability
increases exponentially with the gap $\delta$. 
In other words, the number of sequences folding to a (compact) 
conformation is determined only by the gap associated with the minimum
energy sequence. 

We have shown that the concepts of
designability (i.e. number of sequences folding to a given conformation) and foldability (i.e.,
thermodinamical stability of the sequences with low energy on the given conformation, expressed
by the gap $\delta$) are intimately connected by Eq. (\ref{eq6}). If a protein is stable in its native
conformation, such native conformation is necessarilly highly designable. Vice versa, if a conformation
is highly designable, there exist sequences with a large gap folding to it.

To give a numerical evaluation of protein conformations, we
make use of Eq. (\ref{eq5})) in the form
\begin{equation}
n=\sum_{m=1}^N \frac{k}{m}\exp\left[\left(\alpha-\frac{\overline{\Delta E_2}^2}{4(\sigma_2)^2}\right) m
\right],
\end{equation}
where $k$ does not depend on $m$ and, for the case of the structure displayed in Fig. 1(a),
assumes the value $k=17$ (in keeping with the fact that $\delta=2.5$, $\overline{\Delta
E_2}=1.2$ and $\sigma_2=0.7$).
In the case in which 
$\alpha>\overline{\Delta E_2}^2/4(\sigma_2)^2$, which in the case of the 36mer under
discussion implies $\alpha>0.73$, one can keep only the largest term in the above sum. 
Within this approximation one can write $n\approx e^{1.90\times 36}=0.6\cdot 10^{30}$,
a number to be compared with $n_{36}^{tot}=3.72\cdot 10^{41}$.

One can mention, for the sake of completeness, that the number of sequences within the gap obtained
by pointlike mutations (which is the upper limit to designability), is well fitted by
the function $4\exp(5 m)$, while the total number of sequences is
$19^m {N \choose m}$.

\section{Conclusions}

The
degree of designability of a given conformation depends exponentially
on the energy gap $\delta$.
Since the number of folding sequences is given by the integral of a universal
function (the mutation energy distribution) carried up to $\delta$, a quantity
which  also determines the thermal stability of the designed protein, one can conclude 
that designability and thermal stability are strongly interconnected.
In other words, sequences displaying large
gaps are both thermally stable and
highly designable. Even sequences displaying, in the native conformation,
a small gap 
fold on short call and share (in the compaction process) the conserved contacts leading to
local elementary structures and to the (post--critical) folding nucleus \cite{jchemphys3,aggreg}. 
Consequently, it is
possible to obtain from them, through composition--conserving mutations, 
other sequences
folding to the same native conformation and displaying a large gap. 
In other words, 
any sequence able to fold fast,
folds to a highly designable conformation.

We have estimated that there are of the order of $10^{30}$ sequences folding to a
compact 36mer conformation, over a total of $10^{41}$.
This is only the lower limit, but let us assume that it describes well 
the typical degree of designability of the designed protein. Is this number small or big? The answer to
this question has, of course, important implications from the evolutionary point of view.
If good folders were distributed homogeneously in the space of sequences 
(like in the
case of RNA \cite{schuster}) the important parameter would be their density, that is
$10^{-11}$. This number would be very low, preventing sequences from moving along 
neutral pathways (which are collections of sequences folding to the same conformation
and differing by single mutations). Such a scenario is very unfavourable for evolution.
The situation is however quite different for proteins. 
In fact, it has been shown \cite{isles} that
good folders group themselves in clusters and superclusters, 
giving rise to a quite an inhomogeneous
landscape. Consequently, the relevant
parameter which measures the designability of a conformation 
is the total number of sequences which conserve, 
in any way, the energy gap. This number ($>10^{30}$)
is very large in particular in keeping with the fact that over a life span of the order of 
60 mutations occur  in the genome of each person \cite{creighton}).

\newpage

\begin{figure}
\caption{Three conformations used as natives in the present study. Sequence S${_36}$, which is a good folder onto structure (a), is SQKWLERGATRIADGDLPVNGTYFSCKIMENVHPLA.}
\end{figure}

\begin{figure}
\label{comp}
\caption{Energy distribution for 2 composition--conserving (a) and pointlike (b) mutations.
The parameters of the Gaussian fit (dotted line) are given in the text. }
\end{figure}

\begin{figure} 
\caption{The distribution of energies associated to two composition--conserving
mutations made on three sequences designed on 
three different 36mers conformations. (b) The spectrum obtained
making  two composition--conserving
mutations on three sequences
designed on the same conformation (the one displayed in Fig. 2.4).
The values of the energy gap are $\delta=2.5$ (dotted curve),
$\delta=1.6$ (solid line) and $\delta=1.3$ (dashed line).}
\end{figure}

\begin{figure} 
\caption{The distribution $p(\Delta E)$ associated with two pointlike mutations for the
structure displayed in Fig. 1(a) when the monomers interact with the matrix listed
in Table 5 of ref. \protect\cite{mj} (instead of Table 6). The dashed line is the Gaussian fit
obtained with the least--square method.}
\end{figure}


\bigskip

\begin{thebibliography}{99}
\bibitem{sh_prl} E. I. Shakhnovich, Phys. Rev. Lett. {\bf 72}, 3907 (1994)
\bibitem{sh_gap} V. I. Abkevich, A. M. Gutin and E. I. Shakhnovich, J. Chem. Phys {\bf 101},
6052 (1994)
\bibitem{r3} R. Goldstein, Z. Luthey--Schulten, P. Wolynes, Proc. Natl. Acad. Sci. USA {\bf 89}. 4918
(1992)
\bibitem{r4} A. Sali, E. Shakhnovich and M. Karplus, J. Mol. Biol. {\bf 235}, 1614 (1994)
\bibitem{prl} R. A. Broglia, G. Tiana, H. E. Roman, E. Vigezzi and E. I. Shakhnovich, Phys. Rev. Lett.
{\bf 82}, 4727 (1999)
\bibitem{mj} S. Miyazawa and R. Jernigan, Macromolecules {\bf 18}, 534 (1985)
\bibitem{go} N. Go, Int. J. Peptide Prot. Res. {\bf 7}, 313 (1975)
\bibitem{dill} K. F. Lau and K. Dill, Macromolecules {\bf 22}, 3986 (1989)
\bibitem{ss1} E. I. Shakhnovich, A. M. Gutin, J. Chem. Phys. {\bf 93} 5967 (1989)
\bibitem{sh_b1} E. I. Shakhnovich, Curr. Opin. Struct. Biol. {\bf 7}, 29 (1997)
\bibitem{sh_b2} E. I. Shakhnovich, Folding and Design {\bf 1}, R50 (1996)
\bibitem{bowie} J. Bowie, R. Luthey--Schulten, D. Eisenberg, Science {\bf 253}, 164 (1991)
\bibitem{jchemphys2} G. Tiana and R. A. Broglia, J. Chem. Phys. {\bf 114}, 2503 (2001)
\bibitem{n1} V. I. Abkevich, A. M. Gutin and E. I. Shakhnovich, J. Chem. Phys. {\bf 101}, 6052 (1994)
\bibitem{sh_nucleus} V. Abkevich, A. Gutin and E. Shakhnovich, Biochemistry {\bf 33}, 10026 (1994)
\bibitem{n2} N. Socci, W. Bialek, J. Onuchic, Phys. Rev. E {\bf 49} (1994) 3440
\bibitem{n3} D. Klimov and D. Thirumalai, Phys. Rev. Lett {\bf 76} (1996) 4070
\bibitem{jchemphys1} G. Tiana, R. A. Broglia, H. E. Roman, E. Vigezzi and 
E. I. Shakhnovich, J. Chem. Phys. {\bf 108}, 757 (1998)
\bibitem{jchemphys3} R. A. Broglia and G. Tiana, J. Chem. Phys. (in press)
\bibitem{note_new} Strictly speaking, the Monte Carlo algorithm was to
designed to study equilibrium properties of systems with many degrees of
freedom \protect\cite{metropolis}. Nonetheless, it has been shown
\protect\cite{kikuchi} that, being equivalent to solving the Fokker--Planck
Equation for diffusion in a potential, it can be helpful also 
in studying the kinetical
properties of complex systems, provided that the Fokker--Planck approximation is
valid (i.e., moves are local and the potential changes smoothly on the diffusion
lenght scale). Furthermore, Rey and Skolnick
have shown \protect\cite{dynamicmc} that the folding trajectories obtained with Monte
Carlo simulations are consistent with those obtained with real Molecular
Dynamics calculations.
\bibitem{metropolis} N. Metropolis, A. W. Rosenbluth, M. N. Rosenbluth, A. H.
Teller and E. Teller, J. Chem. Phys. {\bf 21}, 1087 (1953)
\bibitem{kikuchi} K. Kikuchi, M. Yoshida, T. Maekawa and
H. Watanabe, Chem. Phys. Lett. 196, 57 (1992)
\bibitem{dynamicmc} J. Rey and J. Skolnick, Chem. Phys. {\bf 158}, 199 (1991)
\bibitem{note3} This is equivalent to  assume that different degrees 
of optimization are mainly due to "cold" sites, while "hot" sites are always optimized at 
the same degree.
\bibitem{note4} In the case of pointlike mutations this cannot be done, because 
the associated $p_2(\Delta E)$ is
not zero for negative values of the energy, so that energies within the gap can be obtained from
mutations in  "hot" sites, compensated by mutations with negative energy values in other sites.
\bibitem{aggreg} R. A. Broglia, G. Tiana, S. Pasquali, H. E. Roman, E. Vigezzi, Proc. Natl. Acad. Sci. 
USA {\bf 95}, 12930 (1998) 
\bibitem{schuster} P. Schuster and W. Fontana, Physica D {\bf 133}, 427 (1999)
\bibitem{isles} G. Tiana, R. A. Broglia and E. Shakhnovich, Proteins {\bf 39}, 244 (2000)
\bibitem{creighton} T. E. Creighton, {\it Proteins}, J. Freeman and Co., New York (1993)

\end{thebibliography}
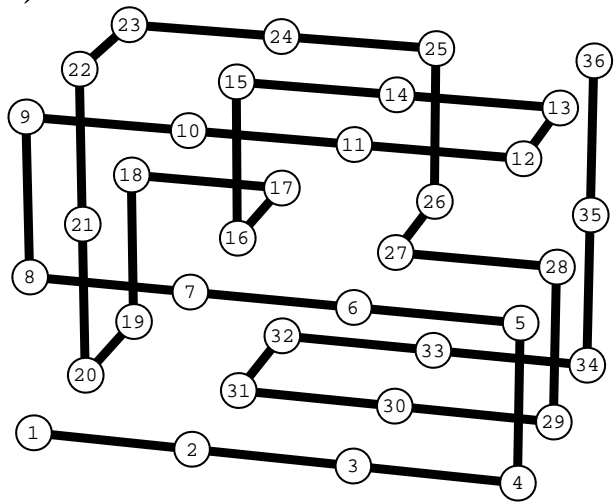
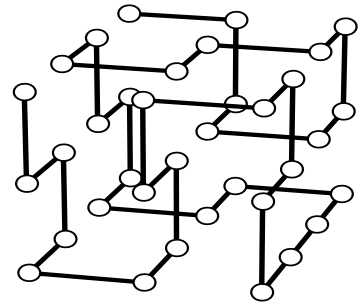
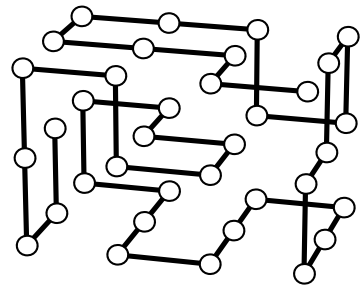
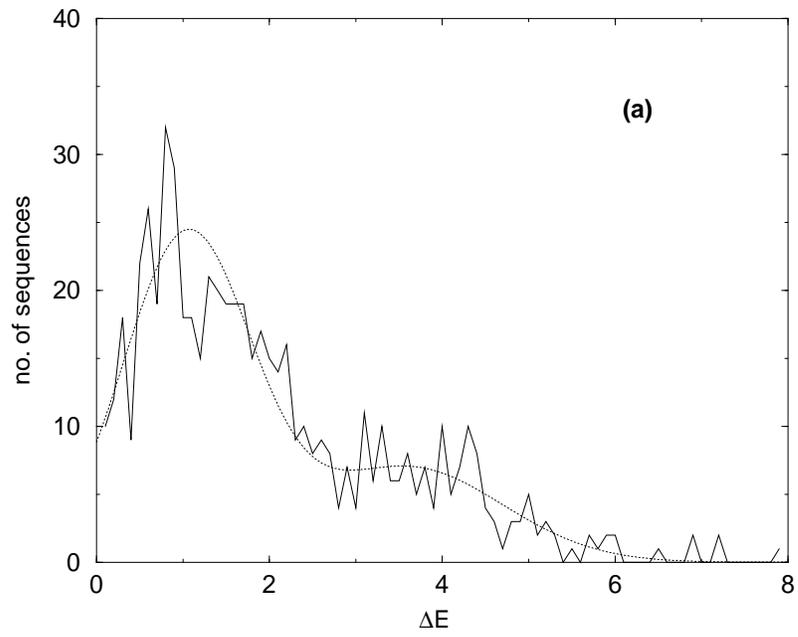
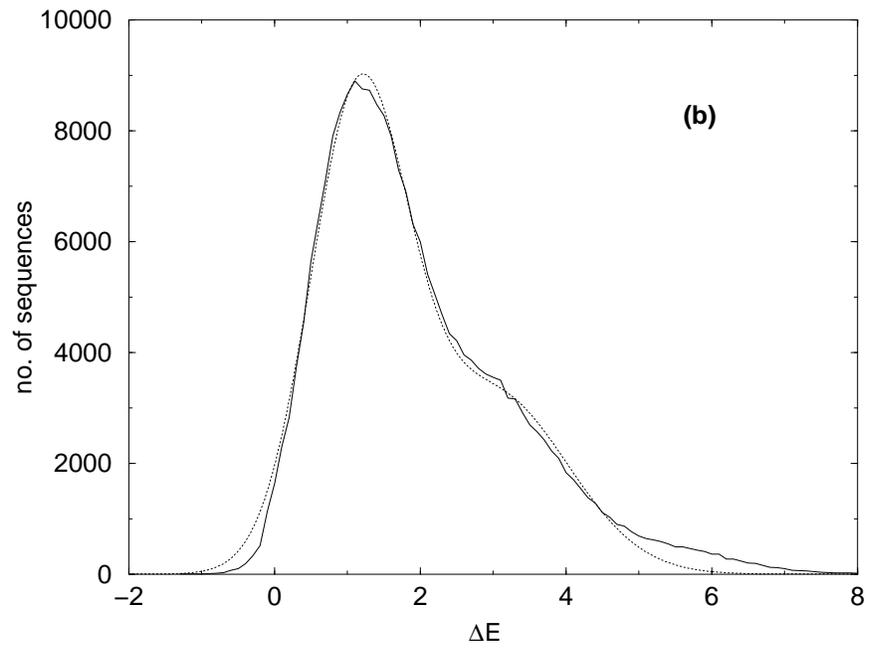
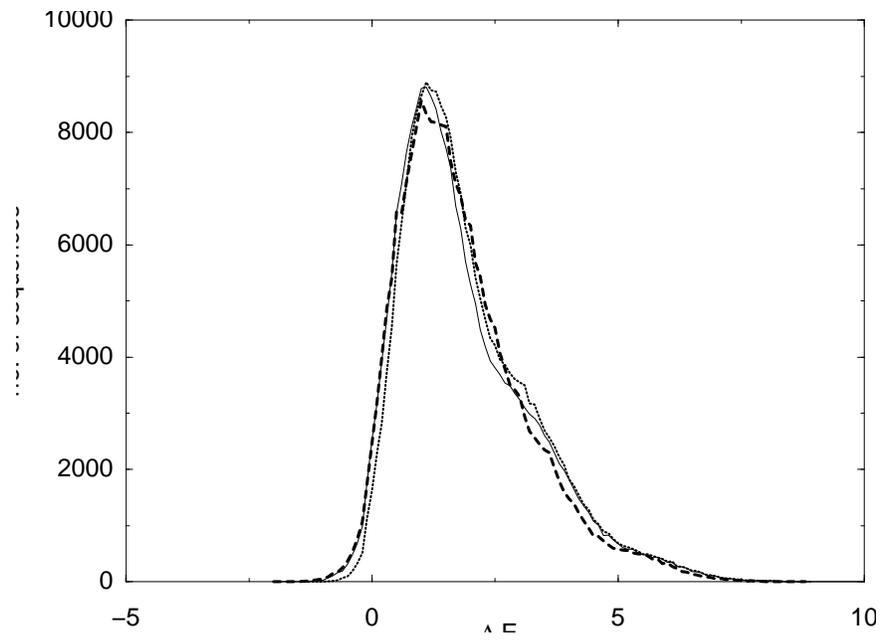
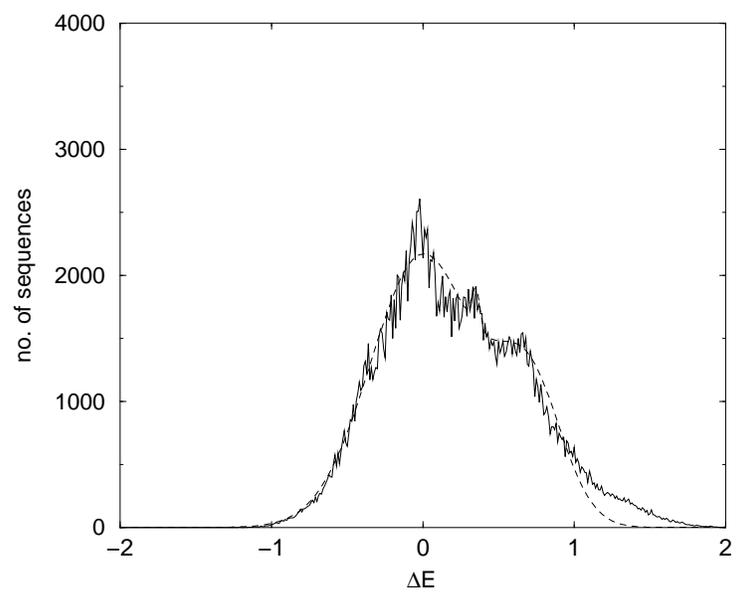
\end{document}